\documentclass{aa}  
\usepackage{graphicx}
\begin{document}

\title{Candidate stellar occultations by large trans-neptunian objects up to 2015\fnmsep\thanks{Tables of predictions for stellar occultations by Eris, Haumea, Ixion, Makemake, Orcus, Quaoar, Sedna, Varuna, 2002~TX$_{300}$, and 2003~AZ$_{84}$ for 2011--2015 and respective Catalog of star positions for 2008--2015 sky path are only available in electronic form at the CDS via anonymous ftp to cdsarc.u-strasbg.fr (130.79.128.5) or via  http://cdsweb.u-strasbg.fr/cgi-bin/qcat?J/A+A/515/A32}
\fnmsep\thanks{Observations made through the ESO run 079.A-9202(A), 075.C-0154, 077.C-0283 and 079.C-0345.}
\fnmsep\thanks{Also based on observations made at Laborat\'orio Nacional de Astrof\'{\i}sica (LNA), Itajub\'a-MG, Brazil.}}

   \titlerunning{Candidate stellar occultations by large TNOs}
   \authorrunning{M. Assafin et al.}

   \author{M. Assafin
          \inst{1}\fnmsep\thanks{Associate researcher at Observatoire de Paris/IMCCE, 77 Avenue Denfert Rochereau 75014 Paris, France},
          J. I. B. Camargo
          \inst{2},
          R. Vieira Martins
          \inst{2,},
          \inst{1}\fnmsep\thanks{Associate researcher at Observatoire de Paris/IMCCE, 77 Avenue Denfert Rochereau 75014 Paris, France}
          F. Braga-Ribas
          \inst{2,},
          \inst{3},
          B. Sicardy
          \inst{3,}
          \inst{4},
          A. H. Andrei
          \inst{2,},
          \inst{1}
          \fnmsep\thanks{Associate researcher at Observatoire de Paris/SYRTE, 77 Avenue Denfert Rochereau 75014 Paris, France}
          \fnmsep\thanks{Associate researcher at INAF/Osservatorio Astronomico di Torino, Strada Osservatorio 20, 10025 Pino Torinese (To), Italy}
          D. N. da Silva Neto
          \inst{5,}
          }

   \offprints{M. Assafin}

   \institute{Observat\'orio do Valongo/UFRJ, Ladeira Pedro Antonio 43,
CEP 20.080-090 Rio de Janeiro - RJ, Brazil\\
              \email{massaf@astro.ufrj.br}
              \and
              Observat\'orio Nacional/MCT, R. General Jos\'e Cristino 77,
CEP 20921-400 Rio de Janeiro - RJ, Brazil\\
              \email{rvm@on.br}
              \and
               Observatoire de Paris-Meudon/LESIA, 5 place Jules Janssen, Meudon, France\\
              \email{bruno.sicardy@obspm.fr}
              \and
               Universit\'e Pierre et Marie Curie, Institut Universitaire de France, Paris, France\\
              \email{bruno.sicardy@obspm.fr}
              \and
              Centro Universit\'ario Estadual da Zona Oeste, Av. Manual Caldeira de Alvarenga 1203,
CEP: 23.070-200 Rio de Janeiro - RJ, Brazil\\
              \email {darionneto@gmail.com}
              }

 \date{Received October 28, 2011; accepted October 29, 2011}

  \abstract
   {We study large trans-neptunian objects (TNOs) using stellar occultations.}
   {We derive precise astrometric predictions for stellar occultations by Eris, Haumea, Ixion, Makemake, Orcus, Quaoar, Sedna, Varuna, 2002~TX$_{300}$, and 2003~AZ$_{84}$ for 2011--2015. We construct local astrometric catalogs of stars complete to magnitudes as faint as R = 18--19 in the UCAC2 (Second US Naval
Observatory CCD Astrograph Catalog) frame covering the sky path of these objects.}
   {During 2007-2009, we carried out an observational program at the ESO2p2/WFI (2.2 m Max-Planck ESO telescope with the Wide Field Imager) instrument. The observations covered the sky path of the selected targets from 2008 to 2015. We performed the astrometry of 316 GB images using the Platform for Reduction of Astronomical Images Automatically (PRAIA). With the help of field distortion patterns derived for the WFI mosaic of CCDs, we reduced the overlapping mosaics of CCDs.}
   {We derive positions in the UCAC2 frame with 40 mas precision for stars up to the catalog magnitude completeness limit (about R = 19). New stellar proper motions are also determined with 2MASS (Two Micron All Sky Survey) and the USNO B1.0 (United States Naval Observatory B 1.0) catalog positions as a first epoch. Astrometric catalogs with proper motions were produced for each TNO, containing more than 5.35 million stars covering the sky paths with 30$\arcmin$ width in declination. The magnitude completeness is about R = 19 with a limit of about R = 21. We predicted 2717 stellar occultation candidates for all targets. Ephemeris offsets with about from 50 mas to 100 mas precision were applied to each TNO orbit to improve the predictions. They were obtained during 2007--2010 from a parallel observational campaign carried out with from 0.6 m to 2.2 m in size telescopes.}
   {This extends our previous work for the Pluto system to large TNOs, using the same observational and astrometric procedures. The obtained astrometric catalogs are useful for follow-up programs at small to large telescopes used to improve the candidate star positions and TNO ephemeris. They also furnish valuable photometric information for the field stars. For each TNO, updates on the ephemeris offsets and candidate star positions (geometric conditions of predictions and finding charts) are made available in the web\thanks{www.lesia.obspm.fr/perso/bruno-sicardy/} by the group.}


   \maketitle
%

\section{Introduction}

   The trans-neptunian objects (TNOs) are located beyond Neptune in a region where only small body differentiation is expected with regard to temperature changes. This makes them 4.5 billion year old interplanetary fossils of the early stages of the formation of the outer Solar System. Recent models of planetary migration such as the Nice model (\cite{LeviMor}; \cite{Rodney2004}; \cite{Rodney2005}) also indicate that TNOs are a sensitive laboratory for the study of orbital dynamics. Refining the ephemeris and establishing the fundamental properties of TNOs are thus essential for our understanding of the origin and evolution of the Solar System. This in turn contributes to the understanding of exoplanetary systems and helps astrochemical and astrobiological studies (\cite{Barucci2008}).

   The analysis of light curves obtained for a stellar occultation observed from many sites allows the shapes and sizes of TNOs to be measured with a precision of a few kilometers. We can probe their atmospheres or place limits on their existence down to the nanobar level. In addition, occultations can eventually lead to the observation or discovery of close binary companions, satellites, or debris around the central body. For size determinations, occultations are far more suitable than indirect estimations such as those coming from albedo assumptions or from the modelling of optical, infrared (IR), and sub-millimetric observations (\cite{Cruikshank}; \cite{Altenhoff};\cite{Stansberry2008}). From these direct size measurements, one can derive more accurate albedos and place tighter constraints on the surface composition of the TNOs. If the mass of the body can be estimated from the orbits of detected satellites or by other indirect means, the density can be estimated more accurately and thus the internal composition and structure of the body can be inferred far more reliably. The detection of atmospheres around TNOs will help to improve our current understanding of their dynamics and relationship with the surface. Detection of Chiron-like jets (\cite{Elliot95}) is also a possibility. A more robust characterization of companions/satellite orbits and rotational periods from stellar occultations and ground-based adaptive optics observations could also improve our models for binary formation and collision (\cite{Goldreich}; \cite{Weidenschilling}; \cite{Sicardy2011a}).

   It is more advisable to select large TNOs as targets for stellar occultation campaigns. They usually have large apparent sizes (30 mas or more) and thus have a greater chance of a positive detection. In addition, theory indicates that larger TNOs are most likely to possess atmospheres (\cite{Elliot2003}). In the case of atmospheres, even negative occultation chords observed nearby locations with positive records are useful, giving valuable upper limits to the atmosphere size and profile. Larger bodies also have smaller relative size errors. The accumulation of positive and negative detections also result in significant improvements in the TNO ephemeris, thus refining the orbit and substantially increasing the accuracy of the subsequent occultation predictions. Improving the chances of success minimizes the money and time spent in the complex international campaigns, which usually involves large telescopes in well-equipped observatories, smaller mobile instruments, and specialized personnel.

   Even as one of the most important and larger members of the TNO population, Pluto has only been probed in terms of occultations, prior to 2000, almost exclusively by the 1988 mutual events with its larger satellite Charon. The first observed stellar occultation by Pluto took place only in 1985, when its atmosphere was first discovered (\cite{Brosch1995}). The atmosphere could only be more extensively observed in the next occultation in 1988 (\cite{Millis1993}). In 2002, a new stellar occultation by Pluto was recorded, but with higher time resolution (\cite{Sicardy2003}; \cite{Elliotal2003}), revealing that there had been a dramatic expansion in the atmosphere pressure. Another high signal-to-noise ratio (S/N) light curve was obtained in the June 2006 stellar occultation (\cite{Young2008}), which uncovered evidence of the stabilization of the atmosphere pressure in Pluto for the period 2002--2008. This result is supported by seven other unpublished occultations observed by our group members in Australia, Brazil, Namibia, and Chile with 0.5 m to 4 m size telescopes in 2006--2010 (Sicardy et al., in prep.). Regarding Charon, after its first recorded stellar occultation (\cite{Walker1980}), four others have so far been observed, one in 11 July 2005 (\cite{Sicardy2006}; \cite{Gulbis2006}), one in 22 June 2008 (\cite{Sicardy2011a}), and two other unpublished multi-chord events observed in Brazil and Chile (4 June 2011) and in Hawaii (23 June 2011). These last two events, as in 22 June 2008, were double events in which Charon and Pluto occulted the same star. In the 4 June 2011 event in particular, both occultations could be recorded at each site. 

   The population of TNOs in general, have been far more poorly studied. Up to the time of writing, only seven stellar occultations have been successfully recorded, all with CCD detectors. A double-chord occultation was observed in Hawaii in October 2009 and involved 2002 TX$_{300}$ and a R = 13.4 UCAC2 star (\cite{Elliot2009}). This was the first time ever that a stellar occultation by a TNO other than Pluto or Charon was recorded. In February 2010, Varuna was observed to occult a V = 11.1 star from the north-east of Brazil. One chord was obtained. A solution for the size, albedo, and shape was derived after combining the positive with a near negative chord and astrometric and photometric ground-based data (Ortiz et al., in prep.). Our group subsequently observed a multi-chord occultation of a R = 16.9 star by Eris in November 2010 (\cite{Maury2010}; \cite{Sicardy2011b}), a one-chord occultation of a R = 18.2 star by 2003 AZ$_{84}$ in January 2011 (\cite{BragaRibas2011a}; Braga-Ribas et al., in prep.), a multi-chord occultation of a R = 18.4 star by Makemake in April 2011 (Ortiz et al., in prep.) and a multi-chord occultation of a R = 15.7 star by Quaoar in May 2011 (Braga-Ribas et al., in prep.). The MIT group also reported at the 218th meeting of the AAS in 2011 the observation of a one-chord occultation by Quaoar in February of 2011 (\cite{Person2011};\cite{Sallum2011}). All seven occultations were predicted in this work. Our group observed them all (except 2002 TX$_{300}$ and the February occultation by Quaoar) with 40 cm to 60 cm size telescopes and also with the NTT and VLT in Chile, Brazil, and Uruguay. 

    The main difficulty in deriving reliable predictions for the stellar occultation of TNOs is their small apparent diameters, even in the case of larger TNOs. Another cause is the lack of accurate orbital elements, resulting in ephemeris errors as large as a few hundreds of milli-arcseconds (mas). Predictions based solely on published catalog positions such as the USNO B1.0 (\cite{usnob1}) or the 2MASS (\cite{2mass}) usually fail because of the poor precision (or lack) of proper motions and the relatively large zonal and/or random errors in their positions (100 mas -- 200 mas). Even individual positions for fainter stars in the UCAC2 catalog (\cite{ucac2}) may need corrections as large as 70 mas.

    There are ways to overcome these ephemeris and star position problems. The ephemeris of TNOs must be offset so as to get realistic TNO positions at predicted occultation dates. The straightforward way to do this is to regularly observe TNOs and derive astrometrically the ephemeris offsets as a function of time. As successive positive occultations are collected, their ephemerides can be radically improved down to a few mas. Even a one-chord positive occultation helps to improve the ephemeris, since the apparent diameters of TNOs are smaller than 30 mas. In the case of star positions, one strategy is to select possible occultations based on the positions given in any arbitrary astrometric catalog, then perform follow-up observations to improve the star position and (after applying offsets to the TNO ephemeris) delineate the shadow path. 

    However, this approach to updating star positions has an intrinsic disadvantage in the case of TNOs. One usually wishes to start searching for occultations from an astrometrically sound catalog in order to select only those truly probable events. The shorter the list of candidates, the greater the chance of conducting the follow-up observational programme on dedicated telescopes in due course. Since the highest quiality astrometric catalogs, such as the UCAC2 for example, are more complete for the brighter magnitudes, then stars fainter than say R = 15 -- 16 will always be left out of the candidate search. This is far from ideal in the case of TNOs. In contrast to Pluto occultations, for TNOs we obtain good contrast in the light curves using very modest instruments for stars as faint as R = 18.0, as verified in the recent successfully recorded TNO occultations reported here. This is due to the TNO magnitudes being much fainter than the star brightness and because of the recent developments in CCD detectors. Choosing the USNO B1.0 or other similar catalogs does not help. The simple addition of fainter stars that do not have the minimum required catalog position precision introduces the problem of enlarging the follow-up list with low quality targets. This wastes valuable telescope time in follow-up programs and may lead to the running out of nights to complete the scheduled objects.
    
    A more suitable strategy for TNOs is to derive local astrometric star catalogs with sufficient position precision, of say 50 mas at least, for the time span of the occultations, for stars in the magnitude range R = 13 -- 19. In this way, we match the required position precision in the search, preserving faint stars without discarding bright objects. The addition of astrometrically trusted faint stars to the follow-up list, avoiding poor quality targets in turn improves the chances of finding more suitable candidates for TNO occultations, owing to the increase in the star density in the sky path.
    
    The first noteworthy consistent studies of Pluto stellar occultation predictions were those of \cite{klemola1985} for the period 1985--1990 and of \cite{McDonald2000a} for the years 1999--2009. Two common drawbacks of these works were the astrometric precision of only about 0$\farcs$2 and the lack of stellar proper motions leading to uncertainties of the order of the Earth radius in the predicted shadow paths. In addition, these earlier predictions were degraded by the poorer precision of the older Pluto ephemerides.

   To overcome the problems faced by these first prediction efforts, we used more modern facilities, namely the most up-to-date CCD detectors, reference catalogs, and astrometric procedures, and carried out an observational program at the ESO2p2/WFI instrument. We derived local star catalogs with precise positions for determining accurate predictions of stellar occultations by Pluto and its satellites Charon, Nix, and Hydra for the period 2008--2015. This work was published in \cite{Assafin2010} and is hereafter referred to as P1. Here, we extend that work to the predict the stellar occultations of the large TNOs Eris, Haumea, Ixion, Makemake, Orcus, Quaoar, Sedna, Varuna, 2002~TX$_{300}$, and 2003~AZ$_{84}$ for 2008--2015, using the same observational and astrometric procedures. The local astrometric catalogs obtained are complete down to about R = 19, have positional errors of 40 mas or smaller, have stars with computed proper motions, and are in the UCAC2 frame. The catalogs cover the TNO sky paths with a $30\arcmin$ width in declination. They can be very useful for the astrometry of small CCD fields for the follow-up of selected candidate stars, and for refining the ephemeris offsets of TNOs for tracking the shadow path across the Earth. The catalogs are also suitable for deriving the photometric properties of calibration and occultation stars.

   In Sect. 2, we describe the ESO2p2/WFI program and observations. The astrometric treatment is given in Sect. 3. In Sect. 4, we present the derived catalogs of star positions along the TNO sky paths. In Sect. 5, we describe the determination of ephemeris offsets and the candidate star search procedure. The predictions of the stellar occultations by the large TNOs are finally presented in Sect. 6. An overview is presented in Sect. 7, including discussions about the external comparison with actual occultation results and error budget estimates, and comments about astrometric follow-up programs for TNO ephemeris improvements.


\section{The ESO2p2/WFI TNO program and observations}

   The TNOs of the ESO2p2/WFI program were selected according to their body size, which is usually related to their apparent sizes on the sky. Given the astrometric uncertainties in the star positions and the error in the TNO ephemeris, larger apparent diameters correspond to a higher probability of a successful prediction. Another selection criteria was the star field. We preferred to select TNOs currently orbiting dense star regions close to the Galactic plane at Galactic longitudes of interest. The TNOs of intrinsically high science value were also selected, even if the star density along their sky path was low. In these cases, the advantage of using the large FOV of the WFI was even clearer.  

   Observations were made at the ESO 2p2 telescope (IAU code 809) using the Wide Field Imager (WFI) CCD mosaic detector. We used the broad-band R filter ESO$\#$844 with $\lambda_{c}$ = 651.725 nm and $\triangle\lambda$ = 162.184 nm (full width at half maximum). The exposure time was 30$^s$. In general, S/N ratios of about 200 were reached for objects of R = 17 without saturating bright (R = 13--15) stars. The limiting magnitude was about R = 21, and the completeness level about R = 19.0. The seeing varied between 0$\farcs$6 and 1$\farcs$5, being typically 1$\arcsec$. The telescope was shifted between exposures in such a way that each star in the path of the TNO was observed at least twice in different CCDs.
   
\begin{figure}
\vspace{0.3cm}
\resizebox{\hsize}{!}{\includegraphics[angle=-90]{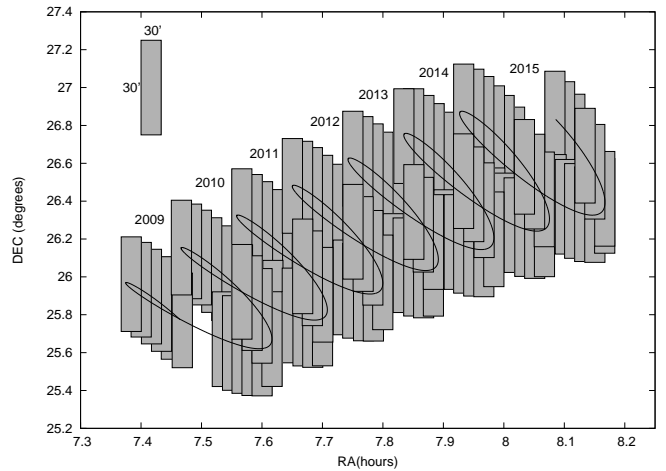}}
\caption{Sky region covered by the ESO2p2/WFI CCD mosaic observations for Varuna. Years 2009--2015 follow from bottom to top. The continuous line is the geocentric sky path of Varuna. Each dashed form represents the 30$\arcmin$ x 30$\arcmin$ area covered by one single WFI mosaic.}
\label{Fig1}%
\end{figure}

   The program was completed in seven runs, in April, June, September, and October 2007, May/June 2008, and February and December 2009. Observations spanned each TNO sky path from 2008 to 2015; in six cases, this path started in 2009, since their WFI observations only took place in 2009 and telescope time was limited. Table~\ref{table:1} shows the approximate limits of the sky paths observed for each TNO, hence indicates the period covered.  Fig.~\ref{Fig1} illustrates the sky path covered by Varuna for 2009--2015. A total of 1167 WFI mosaics or 9336 individual CCD frames were acquired for science, resulting in about 316 GB of photometrically calibrated processed images. The two WFI field distortion patterns derived in P1 were used here. The pattern associated with the September 2007 run was used in all earlier runs; the October 2007 pattern was used in all subsequent runs. The used instruments and observation procedures are described in detail in Sect. 3 of P1.

\begin{table}
\caption{Approximate ($\alpha$,$\delta$) limits and sky path periods covered by ESO2p2/WFI mosaics for each TNO.}
\label{table:1}
\centering
\begin{tabular}{l c c c c c}
\hline\hline
       TNO      &  \multicolumn{2}{c}{($\alpha$,$\delta$)  (ICRF)}  &  \multicolumn{2}{c}{($\alpha$,$\delta$) (ICRF)}   & Time span   \\
                & h ~m & $^{\circ} ~~~~\arcmin$ & h ~m & $^{\circ} ~~~~\arcmin$ & yr \\
\hline                           
Eris            &  01 38 &  $-$04 48 &   01 42 &  $-$02 58 &    2008--2015 \\
Haumea          &  13 40 &  $+$18 57 &   14 01 &  $+$17 32 &    2009--2015 \\
Ixion           &  16 55 &  $-$23 58 &   17 28 &  $-$27 15 &    2009--2015 \\
Makemake        &  12 30 &  $+$28 25 &   12 52 &  $+$25 37 &    2009--2015 \\
Orcus           &  09 40 &  $-$05 09 &   10 04 &  $-$08 42 &    2008--2015 \\
Quaoar          &  17 10 &  $-$15 25 &   17 48 &  $-$15 34 &    2008--2015 \\
Sedna           &  03 27 &  $+$06 28 &   03 41 &  $+$07 17 &    2009--2015 \\
Varuna          &  07 30 &  $+$25 45 &   08 03 &  $+$26 35 &    2009--2015 \\
2002~TX$_{300}$ &  00 32 &  $+$26 59 &   01 07 &  $+$32 43 &    2008--2015 \\
2003~AZ$_{84}$  &  07 36 &  $+$12 17 &   08 02 &  $+$09 58 &    2009--2015 \\
\hline
\end{tabular}
\begin{list}{}
\item  Approximate geocentric ($\alpha$,$\delta$) limits and corresponding sky paths periods covered for each TNO. Paths 2008--2010 were observed in the September and October 2007 runs, and 2009--2015 ones were observed in the 2009 runs at ESO2p2. Fig.~\ref{Fig1} illustrates the actual sky path for Varuna.
\end{list}
\end{table}

\section{Astrometry}

   We briefly describe the reduction procedures. All CCD images were corrected by overscan, zeromean, flatfield, and bad pixels with IRAF (\cite{Tody1993}) using the esowfi (\cite{Jones2000}) and mscred (\cite{Valdes1998}) packages. Using the PRAIA package (\cite{PRAIA}), the astrometric treatment for each TNO consisted of three distinct steps. 

\par

An assigned field distortion pattern was first applied to each CCD in the WFI mosaics for each run (see details in Sect. 4.1 of P1).

\par

\begin{figure}
\hspace{-0.5cm}
\vspace{-0.7cm}
\sidecaption
\includegraphics[width=6.5cm,angle=-00]{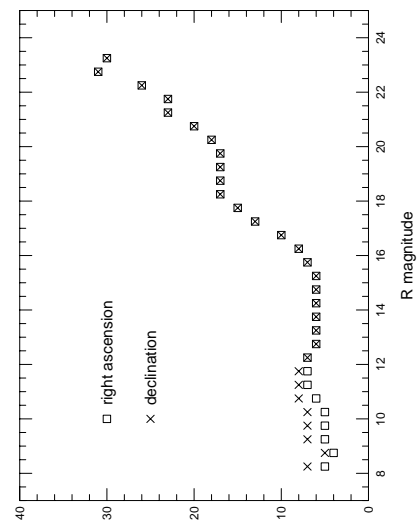}
\caption{(x, y) measurement errors as a function of R magnitude from all treated CCDs. Values are averages in 0.5 magnitude bins.}
\label{Fig2}%
\end{figure}

In the second step, astrometry was performed over the individual CCDs, where the (x, y) measurements were corrected by the pre-determined field distortions. Magnitudes were obtained from PSF photometry and calibrated with respect to the UCAC2. The (x, y) measurements were performed with two--dimensional circularly symmetric Gaussian fits. The estimated (x, y) measurement errors from Gaussian fits were about 5 -- 10 mas for R $<$ 16, then increased to 18 mas for R $<$ 18, to 20 mas for 18 $<$ R $<$ 20, and to 30 mas at fainter magnitudes. Fig.~\ref{Fig2} shows the distribution of (x, y) errors as a function of R magnitude for all measurements of all TNOs. The UCAC2 was the prime reference catalog, except for Eris. In this case, the 2MASS catalogue was used as the reference frame owing to the insufficient number of UCAC2 stars in the FOV. This is not an important change, as in the last step, all frames were consistently corrected to the UCAC2 frame. In steps 1 and 2, we used the six constant polynomial model to relate the (x, y) measurements to the (X, Y) tangent plane coordinates. The position mean errors in the ($\alpha$,$\delta$) individual CCD solutions are listed in Table~\ref{table:2}; the average number of reference stars and number of CCD frames are also given for each TNO set of observations. Detailed information on the procedures in this step are found in Sect. 4.2 of P1.

\begin{table}
\caption{Astrometry of individual CCD frames of WFI mosaics.}
\label{table:2}
\centering
\begin{tabular}{l c c c c}
\hline\hline
                  & \multicolumn{2}{c}{Mean errors}                                 & No. of          &  No. ref.       \\
TNOs              & $\sigma(\Delta$$\alpha$cos$\delta)$  & $\sigma(\Delta$$\delta)$ & frames          &  stars/frame    \\
                  &      mas                             &     mas                  &                 &                 \\
\hline
Eris              &       97                             &      95                  &       576       &           27    \\
Haumea            &       51                             &      50                  &       784       &            8    \\
Ixion             &       54                             &      53                  &       976       &          108    \\
Makemake          &       48                             &      51                  &       800       &            7    \\
Orcus             &       58                             &      55                  &      1528       &           14    \\
Quaoar            &       57                             &      57                  &       904       &          121    \\
Sedna             &       54                             &      57                  &       576       &           11    \\
Varuna            &       49                             &      53                  &       848       &           31    \\
2002~TX$_{300}$   &       47                             &      50                  &      1184       &           20    \\
2003~AZ$_{84}$    &       50                             &      50                  &      1160       &           48    \\
\hline
\end{tabular}
\begin{list}{}
\item Mean errors are the standard deviations in the (O$-$C) residuals from individual ($\alpha$,$\delta$) solutions with the UCAC2 catalog, except for Eris in which case the 2MASS catalogue was used as the reference frame (see text for details). The six constant polynomial model was used to relate the (x, y) measurements to the (X, Y) tangent plane coordinates. The (x, y) centers were pre-corrected by the field distortion patterns (see details of these pre-corrections in Sect. 4.1 of P1).
\end{list}
\end{table}

\par

In the final step, all positions coming from the different CCDs and mosaics were combined to produce a global solution for each sky path year, and final ($\alpha$,$\delta$) star positions were obtained in the UCAC2 system. In this step, all positions were projected onto the tangent plane and a complete polynomial model of the third degree was used to fix all the coordinates in the UCAC2 frame. Table~\ref{table:3} lists the mean errors of UCAC2 stars before and after the WFI mosaic global solutions. The listed values are averages taken over all sky path years. In addition to eliminating zero-point errors present in the positions in the individual CCD frames, the overall standard deviations with respect to the UCAC2 catalogue were generally smaller, as can be seen in Table~\ref{table:3}. Detailed information about the global reduction procedures are given in Sect. 4.3 of P1.

\begin{table}
\caption{Global astrometric solution for WFI CCD mosaics.}
\label{table:3}
\centering
\begin{tabular}{l c c c c c}
\hline\hline
       & \multicolumn{4}{c}{($\alpha$,$\delta$)--UCAC2}  & UCAC2~\\
       & \multicolumn{2}{c}{ before G.S.}  &  \multicolumn{2}{c}{ after G.S. }       &   Star  \\
TNOs   &   $\sigma_{\alpha}$ &  $\sigma_{\delta} $ & $\sigma_{\alpha}$ &  $\sigma_{\delta}$ & No. \\
       &      mas            &       mas           &      mas          &    mas      &        \\
\hline
Eris             &         62             &     63   &        49     &    54  &        232    \\
Haumea           &         45             &     44   &        46     &    46  &        617    \\
Ixion            &         66             &     65   &        53     &    52  &       6229    \\
Makemake         &         41             &     43   &        46     &    49  &        523    \\
Orcus            &         80             &     73   &        56     &    54  &        842    \\
Quaoar           &         96             &     91   &        56     &    55  &       5996    \\
Sedna            &         50             &     53   &        53     &    53  &        396    \\
Varuna           &         53             &     57   &        49     &    52  &       1793    \\
2002~TX$_{300}$  &         58             &     64   &        46     &    50  &       1617    \\
2003~AZ$_{84}$   &         56             &     56   &        49     &    49  &       3706    \\
\hline
\end{tabular}
\begin{list}{}
\item The ($\sigma_{\alpha}$, $\sigma_{\delta}$) refer to the observed minus UCAC2 positions computed before and after the global solution (G.S.) procedure. Last column is the number of used UCAC2 reference stars. All values are averages over the results for each sky path year. For more details about the mosaic global solution, see Sect. 4.3 of P1. 
\end{list}
\end{table}

In this final step, multiple entries coming from distinct CCD frames for the same objects are resolved (see Sect. 4.4 of P1 for details). Table~\ref{table:4} displays the statistics for resolved multiple entries. The listed values are taken over the global solutions from all sky path years. Each final object position is assigned a flag that indicates its multiplicity status (see Table~\ref{table:4} for details). In all, more than 93 \% of the final positions come with no multiple entries (flag = 0).

\begin{table}
\caption{Multiplicity flags for WFI global mosaic positions.}
\label{table:4}
\centering
\begin{tabular}{l c c c c c c}
\hline\hline
TNOs              &   f0   &   f1   &   f2   &   f3   &    f4  &  f5   \\
                  &   \%   &   \%   &   \%   &   \%   &    \%  &  \%   \\
\hline
Eris              &   86.1  &   4.3  &  1.3   &  2.8  &  5.4  & 0.1  \\
Haumea            &   93.1  &   0.9  &  0.4   &  0.7  &  4.8  & 0.1  \\  
Ixion             &   93.8  &   0.3  &  1.3   &  2.6  &  2.0  & 0.0  \\             
Makemake          &   90.3  &   1.9  &  0.6   &  1.2  &  5.9  & 0.1  \\              
Orcus             &   82.3  &   9.1  &  3.2   &  1.9  &  3.5  & 0.0  \\              
Quaoar            &   92.9  &   1.8  &  1.0   &  1.7  &  2.6  & 0.0  \\               
Sedna             &   92.3  &   3.0  &  0.9   &  1.0  &  2.8  & 0.0  \\               
Varuna            &   93.9  &   1.0  &  0.5   &  1.2  &  3.4  & 0.0  \\               
2002~TX$_{300}$   &   94.2  &   2.0  &  1.0   &  0.4  &  2.3  & 0.1  \\               
2003~AZ$_{84}$    &   95.6  &   0.3  &  0.2   &  1.1  &  2.7  & 0.1  \\               
All TNOs          &   93.5  &   0.8  &  1.1   &  2.2  &  2.3  & 0.1  \\               
\hline
\end{tabular}
\begin{list}{}
\item Multiplicity flag statistics for each and all TNOs. Percentages refer to the total number of mosaic positions for each TNO and for all TNOs together (last line). No flag (good astrometry) means no multiple entries within 1$\farcs$5 for the same object after mosaic solutions. Flag cases f1 and f2 apply only to UCAC2 and 2MASS stars, when only entries assigned to a catalog were used. Flag f1 means that more than one entry was used (more than one position entry was averaged), whereas flag f2 indicates that only one entry was used. Flags f3, f4, and f5 apply only to field stars and mean that only one entry was selected according to one of three criteria, in order of priority: a) highest number of used common individual CCD positions (flag f3); b) least (x, y) measurement error (flag f4); c) brightest R magnitude (flag f5). 
\end{list}
\end{table}

Proper motions were extracted directly from the UCAC2 catalog. For the other stars, proper motions were computed for each star using the 2MASS and USNO B1.0 catalogs as a first epoch. Although these two catalogs may have random/systematic errors at the 100--200 mas level, they are still the most suitable choice for this purpose. The proportion of UCAC2, 2MASS-based, and USNOB1-based proper motions of stars are given in Table~\ref{table:5}. Details about the computations can be found in Sect. 4.5 of P1. The highest percentage of stars without proper motions was verified for Ixion, namely 62 \%, which is twice the average fraction (about 30\%). Ixion is now crossing the plane of the Milky Way (see Table~\ref{table:1}). One possible explanation is that the WFI magnitude limit is fainter than that of the USNO B1.0 (see star distribution according to magnitude in Fig.~\ref{Fig3} next in the text) and many faint stars are present in this region of the sky.

\begin{table}
\caption{Proper motion computations from 2MASS, USNO B1.0, and ESO2p2/WFI global mosaic star positions.}
\label{table:5}
\centering
\begin{tabular}{l c c c c}
\hline\hline
                  &   UCAC2   &  2MASS     & USNO B1.0   &  No   \\
TNOs              &   p.m.    &  p.m.      & p.m.     &  p.m.    \\
                  &   \%      &   \%       &   \%     &   \%     \\
\hline
Eris              &   8.4  &   35.5  &  32.9   &   23.2  \\
Haumea            &   6.3  &   24.2  &  41.6   &   27.9  \\  
Ixion             &   1.7  &   22.3  &  13.5   &   62.5  \\             
Makemake          &   6.4  &   25.2  &  38.8   &   29.6  \\              
Orcus             &  15.6  &   50.3  &  22.1   &   12.0  \\              
Quaoar            &   5.9  &   41.3  &  21.6   &   31.2  \\               
Sedna             &  12.2  &   49.0  &  22.8   &   16.0  \\               
Varuna            &  10.7  &   34.3  &  34.7   &   20.3  \\               
2002~TX$_{300}$   &  18.2  &   48.0  &  24.6   &   09.2  \\               
2003~AZ$_{84}$    &  10.9  &   27.8  &  34.5   &   26.8  \\               
\hline
\end{tabular}
\begin{list}{}
\item UCAC2 stars with proper motions (directly extracted from UCAC2), stars with proper motions based on the 2MASS and USNO B1.0, and stars for which no proper motion could be computed. Percentages are relative to the total number of stars for each TNO.
\end{list}
\end{table}

 A complete description of the astrometric and photometric procedures used here can be found in Sect. 4 of P1, where data for the same WFI CCD mosaic underwent rigorously the same treatment.

\section{The catalogs of star positions of each TNO for 2008--2015 sky paths}

For each TNO, we produced star catalogs following their respective sky paths up to 2015. The paths begin in 2008 for Eris, Orcus, Quaoar, and 2002~TX$_{300}$, and in 2009 for the other TNOs. The star catalogs consist of the mean ($\alpha$,$\delta$) positions in the ICRS (J2000), proper motions, R magnitudes (also J, H, and K in the case of 2MASS stars), mean epoch of observations, position error at the mean epoch of observation, and the magnitude error estimates. In all, 5,356,255 stars were catalogued in the UCAC2 frame. The mean epoch is approximately 2007.75 for the catalogs of TNOs beginning in 2008, and about 2009.5 for the others. Magnitude completeness is about R = 19. The magnitude limit is about R = 21. Position error is smaller than 40 mas for stars up to magnitude R = 19, and 30 mas up to R = 17. The catalogs are available in electronic form at the CDS.

The catalogs are divided in terms of year. Stars that had multiple entries within 1$\farcs$5 in the global mosaic solutions are flagged (see Table~\ref{table:4}). The R magnitudes from PSF photometry were calibrated in the UCAC2 system, thus magnitude zero-point errors up to 0.3 are expected for R $>$ 17. The position error is estimated from repeated observations, by evaluating the standard deviation (mean error) in the contributing individual CCD positions about the final catalog star positions (last step of global mosaic solution - see Sect. 3). By default, multiple-entry flagged stars have no position error estimates. Infrared magnitudes (and errors) were extracted from the 2MASS catalog. Error estimates for R magnitudes come from the standard deviation about the mean for individual CCD frames. The sky coverages of the catalogs have widths of at least $30\arcmin$ in declination and lenghts of a few degrees along the geocentric sky path of each TNO. The proportion of UCAC2, 2MASS, and USNO B1.0 stars in the WFI catalogs may be verified from Table~\ref{table:5}.

\begin{figure}
\centering
\hspace{+0.2cm}
\vspace{-0.5cm}
\sidecaption
\includegraphics[width=6.3cm,angle=-00]{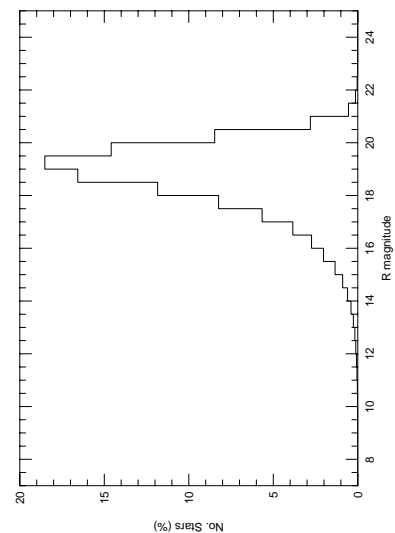}
\caption{Star distribution per R magnitude for all catalog stars. It illustrates the R magnitude completeness and limit of the catalogs. Percentages were computed over 0.5 magnitude bins for all stars in all catalogs.}
\label{Fig3}%
\end{figure}

\begin{figure}
\centering
\hspace{-0.2cm}
\vspace{-0.5cm}
\sidecaption
\includegraphics[width=6.3cm,angle=-00]{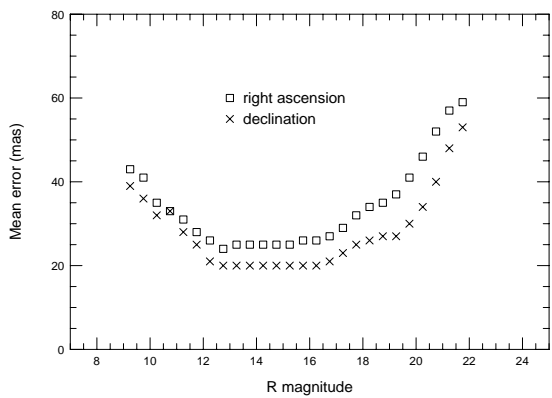}
\caption{Position mean errors as a function of R magnitude for all catalog stars. Position errors are estimated from the standard deviation in the contributing individual CCD positions about the final catalog star positions (last step of global mosaic solution - see Sect. 3). Values were computed over 0.5 magnitude bins.}
\label{Fig4}%
\end{figure}

For each TNO, Table~\ref{table:6} lists the total number of catalog stars per year, the R bandpass magnitude completeness, magnitude limit, and average position errors for faint stars with R at the magnitude completness. Fig.~\ref{Fig3} shows the star distribution per R magnitude for all catalogs. Fig.~\ref{Fig4} plots the position error of all catalog stars as a function of R magnitude. Values were averaged over 0.5 magnitude bins.

\begin{table}
\caption{Star catalogs for each TNO sky path for 2008/2009--2015.}
\label{table:6}
\centering
\begin{tabular}{l c c c c c}
\hline\hline
       TNO        & No. stars   &   $\sigma_{\alpha}$       &   $\sigma_{\delta}$         &   R magnitude           &   R magnitude        \\
                  &             &   mas                     &      mas                    &   completeness          &    limit             \\
\hline
Eris              &  25,685      &   33  &  37   &   19.0 &   21.0 \\
Haumea            &  77,860      &   41  &  36   &   19.5 &   21.5 \\  
Ixion             &  3,452,835   &   41  &  29   &   19.5 &   21.5 \\             
Makemake          &  64,487      &   42  &  39   &   19.5 &   21.5 \\              
Orcus             &  56,042      &   40  &  39   &   18.5 &   20.5 \\              
Quaoar            &  1,150,375   &   41  &  37   &   18.5 &   20.5 \\               
Sedna             &  27,464      &   33  &  31   &   19.0 &   21.0 \\               
Varuna            &  140,041     &   38  &  39   &   19.0 &   21.0 \\               
2002~TX$_{300}$   &  82,746      &   33  &  37   &   18.0 &   20.0 \\               
2003~AZ$_{84}$    &  278,720     &   33  &  34   &   19.0 &   21.5 \\               
\hline
\end{tabular}
\begin{list}{}
\item Number of catalog stars per TNO, position error for faint stars with R at completeness magnitude (errors are estimated from the standard deviation of contributing individual CCD positions about the final catalog star positions), R bandpass magnitude completeness, and magnitude limit.
\end{list}
\end{table}

\section{Ephemeris offsets and search for candidate stellar occultations}

To achieve the highest accuracy in predicting stellar occultations, it is necessary to pin down the actual paths of the TNOs in the sky. This can be done by correcting their orbits by offseting their ephemerides in both right ascension and declination. We derived average ephemeris offsets for all TNOs in the ESO2p2/WFI program. Table~\ref{table:7} displays the ephemeris offsets information for each TNO. The TNO observed positions used to derive the offsets were in the UCAC2 system. They came from a parallel astrometric program of CCD observations carried out during 2007--2010 at a variety of 0.6 m to 2.2 m telescopes in Chile (ESO2p2/WFI), Brazil (Laborat\'orio Nacional de Astrof\'{\i}sica - LNA), and France (Pic du Midi; Observatoire Haute de Provence). The obtained offsets are provisional in the sense that the aforementioned program is still ongoing and further refinements and updates are expected. For three TNOs (Varuna, 2002~TX$_{300}$, 2003~AZ$_{84}$), we derived the offsets from fits to actual observed stellar occultations (see discussion in next Sect. 7). The offset error estimates come from the standard deviation about the mean of individual nightly averages. They do not depend upon the particular ephemeris errors themselves, which were published by the JPL.

\begin{table}
\caption{TNOs offsets with respect to JPL ephemerides.}
\label{table:7}
\centering
\begin{tabular}{c c c c c c c}
\hline\hline
              & \multicolumn{4}{c}{Observed - JPL ephemeris} & No. & JPL  \\
              &  $\Delta\alpha$cos$\delta$ &  $\Delta\delta$ & E$_{\alpha}$ & E$_{\delta}$ & Nights & ephemeris  \\
              &  mas   &  mas  & mas & mas & & version  \\
\hline                           
Eris           &  $-$020 & $-$068 &  084  &  055 &  09  & 28  \\
Haumea         &  $-$090 & $-$098 &  060  &  081 &  21  & 37  \\
Ixion          &  $+$020 & $+$191 &  092  &  093 &  15  & 09  \\
Makemake       &  $-$039 & $-$165 &  060  &  058 &  20  & 40  \\
Orcus          &  $-$087 & $+$001 &  059  &  046 &  19  & 13  \\
Quaoar         &  $-$144 & $-$078 &  098  &  043 &  22  & 17  \\
Sedna          &  $-$043 & $-$160 &  040  &  080 &  05  & 17  \\
Varuna         &  $-$045 & $+$005 &  015  &  015 &  --  & 21  \\
2002~TX$_{300}$&  $-$084 & $-$053 &  040  &  040 &  --  & 14  \\
2003~AZ$_{84}$ &  $+$232 & $+$006 &  003  &  015 &  --  & 11  \\
\hline
\end{tabular}
\begin{list}{}
\item  Average $"$observed minus JPL ephemeris$"$ position offsets. Errors come from the standard deviation about the mean of individual nightly averages. TNO positions in the UCAC2 system come from a parallel astrometric program of CCD observations carried out during 2007--2010 at a variety of 0.6 m to 2.2 m telescopes in Chile, Brazil, and France. Each TNO ephemeris version is displayed. They were extracted using the JPL Horizons service (\cite{Giorgini1996}). The listed offsets are provisional in the sense that the parallel program is still ongoing and further updates are expected. For three TNOs (Varuna, 2002~TX$_{300}$, 2003~AZ$_{84}$), offsets came from fits to actual observed stellar occultations (see discussion in Sect. 7).   
\end{list}
\end{table}

   Using the astrometric catalog of stars in the path of the sky through which the TNOs will pass up to 2015, we then searched for stars that could be occulted by the respective TNOs. To achieve this, the whole star catalogs were cross-correlated with the JPL ephemerides of the bodies generated each minute for the whole period between 2008/2009 to 2015. The JPL ephemerides were offset prior to the search using the quantities given in Table~\ref{table:7}. If the apparent distance in the sky plane between the star and the body is less than an Earth apparent radius plus 50 mas of safe margin, then a potential occultation was identified and all astrometric and geometric data relevant to the possible event were computed and stored. In addition to astrometric and photometric data, for each candidate star, we computed and stored the minimum apparent geocentric distance $d$, the central instant of closest approach $t_{0}$, the shadow velocity $v$ across the Earth, the position angle $P/A$ of the shadow path, and the local solar time $LST$ at the sub-planet point. These geometric quantities are defined and calculated as explained in Sect. 7 of P1.

\section{Predictions of stellar occultations by ten large TNOs}

Following the procedure described in Sect. 5, we identified potential candidate stars for occultations by Eris, Haumea, Ixion, Makemake, Orcus, Quaoar, Sedna, Varuna, 2002~TX$_{300}$, and 2003~AZ$_{84}$. The adopted search radius was  0$\farcs$335 - i.e. 285 mas (approximately the apparent Earth radius at 31 AU, i.e., at Pluto for 2008--2015) plus a safe margin of 50 mas. No predictions were discarded due to day light at the sub-planet point, as occultations could, even so, be visible right above the horizon from places still in the dark near the Earth terminator. For each TNO, all relevant astrometric, photometric, and geometric information for each potential event found is available in electronic form\footnote{anonymous ftp to cdsarc.u-strasbg.fr}. Table~\ref{table:8} lists a sample of predictions for 2002~TX$_{300}$, Varuna, Eris, 2003~AZ$_{84}$, Makemake, and Quaoar. These very predictions were used to prepare campaigns actually resulting in successful recordings of stellar occultations by our group. The exception was 2002~TX$_{300}$, for which no observations could be done at Hawaii (the predicted path), because all telescopes were occupied, must of them having been scheduled to observe the impact of NASA's Lunar Crater Observation and Sensing Satellite (LCROSS; \cite{LCROSS}), which occurred about one hour after the predicted central instant of occultation. Prediction tables contain the date and instant of stellar occultation (UTC), both the ICRS (J2000) star and TNO estimated coordinates at the event date, the closest apparent geocentric distance between the star and TNO, the position angle of the shadow across the Earth (clockwise, zero at North), the velocity in km s$^{-1}$, the distance to the Earth (AU), longitude of the sub-solar point, local solar time, applied JPL ephemerides offsets in (RA, DEC) for the central instant, catalog proper motion and multiplicity flags, estimated star catalog position errors, proper motions, and magnitudes R*, J*, H*, and K*. Magnitudes were normalized to a reference shadow velocity of 20 km s$^{-1}$ by

\begin{displaymath}
M^{*} = M + 2.5~log_{10} \left(\frac{v}{20 ~ km ~ s^{-1}}\right)
\end{displaymath}

The reference velocity of 20 km s$^{-1}$ is typical of events around Pluto opposition. In slow events, M$^*$ may become one to two magnitudes brighter than the actual star magnitude, indicating that the observation is feasible after all, because we can use longer integration times, and consequently reach reasonably high S/N, without any losses of spatial resolution in diameter measurements. For the same reasoning, we may also probe the atmosphere altitudes in the light curves, despite the faintnesses of the stars. As mentioned in the introduction, in the case of TNOs, occultations involving stars as faint as R = 18.0 have been successfully recorded with 50 cm telescopes.

\begin{table*}
\caption{Sample from prediction tables for stellar occultations by large TNOs.}
\label{table:8}
\centering
\begin{tabular}{c c c c c}
\hline\hline
  year m~ d  ~~h ~m s     &  RA ~~(ICRS)~~ Dec &  C/A  P/A  ~~~~  v  ~~~ D ~~&  R* ~  J* ~  H*  ~  K* & $\lambda$ ~ LST  ~~ $\Delta$e$_{\alpha}$ ~ $\Delta$e$_{\delta}$  pm ct fg ~~ E$_{\alpha}$ ~ E$_{\delta}$ ~ $\mu_{\alpha}$ $\mu_{\delta}$  \\
                       & h ~m ~~~s ~~~~~~~$^{\circ}$ ~~~$\arcmin$ ~~~~~$\arcsec$ & mas ~  $^{\circ}$ ~~~~~  km s$^{-1}$   AU~ &   & ~  $^{\circ}$ ~  h:m  ~~  mas ~~ mas ~~~~~~~~ ~~~ mas mas mas mas \\
\hline                           
 2009 10 09 ~10 29 29  & 00 37 13.6205 +28 22  23.019 & 034  158.89  ~-25.29~  40.61 & 13.5 12.6 12.2 12.2 & 194  23:24  -84.0  -53.0  ok  uc  0 ~~ 11 ~ 04  $-$16  $-$11 \\
 2010 02 19 ~23 07 40  & 07 29 22.4672 +26 07  23.191 & 031  188.28  ~-18.66~  42.75 & 11.8 10.3 10.0 10.0 & 336  21:30  -45.0  +05.0  ok  uc  0 ~~ 40 ~ 70  $-$01  $-$12 \\
 2010 11 06 ~02 08 11  & 01 39 09.9421 -04 21  12.119 & 021  347.56  ~-26.32~  95.74 & 16.8 14.3 13.7 13.5 & 308  22:38  -20.0  -68.0  ok  2m  0 ~~ 06 ~ 19  $+$90  $-$12 \\
 2011 01 08 ~06 31 48  & 07 43 41.8220 +11 30  23.569 & 094  187.02  ~-26.00~  44.32 & 18.5 50.0 50.0 50.0 & 271  00:33 +232.0  +06.0  ok  fs  0 ~~ 11 ~ 21  $-$03  $-$03 \\
 2011 04 23 ~01 37 59  & 12 36 11.3938 +28 11  10.493 & 174  183.80  ~-22.09~  51.50 & 18.4 50.0 50.0 50.0 & 314  22:33  -39.0 -165.0  ok  fs  0 ~~ 18 ~ 19  $-$05  $-$10 \\
 2011 05 04 ~02 40 45  & 17 28 50.8021 -15 27  42.788 & 046  191.77  ~-18.28~  42.35 & 15.7 13.5 12.8 12.7 & 000  02:42 -144.0  -78.0  ok  uc  0 ~~ 36   13  $-$04  $-$11 \\
\hline
\end{tabular}
\begin{list}{}
\item In entry order, the above samples correspond to the predictions for 2002~TX$_{300}$ (October 2009), Varuna (February 2010), Eris (November 2010), 2003~AZ$_{84}$ (January 2011), Makemake (April 2011), and Quaoar (May 2011), respectively. Prediction tables list event date and instant (UTC), the ICRS (J2000) star coordinates at occultation, the closest apparent distance between star and TNO (C/A), the position angle (P/A) of the shadow across the Earth (counter-clockwise, zero at north), the velocity in km s$^{-1}$, the distance (D) to the Earth (AU), R*, J*, H*, and K* star magnitudes normalized to a reference shadow velocity ($v$) of 20 km s$^{-1}$ (50.0 means no magnitude available), longitude ($\lambda$) of the sub-solar point, local solar time (LST), ($\Delta$e$_{\alpha}$, $\Delta$e$_{\delta}$) JPL ephemeris offsets in ($\alpha$,$\delta$) for the central instant (see Sect. 5), catalog cross-identification (uc = UCAC2, 2m = 2MASS, fs = field star), proper motion existence and multiplicity flags (see Sect. 3), estimated star-catalog position errors (E$_{\alpha}$, E$_{\delta}$), and proper motions ($\mu_{\alpha}$, $\mu_{\delta}$). Positive/negative $v$ means, respectively, prograde/retrograde velocities, that is TNO's geocentric right ascension is increasing/decreasing, respectively. The complete table sets of 2008/2009--2015 predictions for Eris, Haumea, Ixion, Makemake, Orcus, Quaoar, Sedna, Varuna, 2002~TX$_{300}$, and 2003~AZ$_{84}$ are available in electronic form at the CDS. In the electronic version, in addition to the star positions, we provide the TNO estimated coordinates (with ephemeris offsets applied) at occultation instant.
\end{list}
\end{table*}

Table~\ref{table:9} displays the total number of predicted events for each TNO for the period 2008/2009--2015.

\begin{table}
\caption{Predictions of stellar occultations by ten large TNOs for 2008--2015.}
\label{table:9}
\centering
\begin{tabular}{l c c c c c c c c}
\hline\hline
 TNO           & 2008     &     2009     &     2010    &    2011     &     2012    &      2013    &     2014    &     2015 \\
\hline
Eris           &    2     &        1   &       1     &         0   &         0    &         1    &         0    &        0 \\
Haumea         &    x     &        3   &       4     &         6   &         4    &         7    &         7    &        6 \\
Ixion          &    x     &      102   &     193     &       288   &       352    &       425    &       309    &      147 \\
Makemake       &    x     &        4   &       3     &         5   &         3    &         1    &         4    &        0 \\
Orcus          &    4     &        3   &       6     &         1   &         3    &         1    &         2    &        0 \\
Quaoar         &   51     &       26   &      27     &        90   &       176    &        53    &        67    &      117 \\
Sedna          &    x     &        1   &       0     &         2   &         1    &         2    &         3    &        3 \\
Varuna         &    x     &        9   &      11     &        11   &         8    &        14    &         7    &        8 \\
2002~TX$_{300}$&    1     &        4   &       3     &         3   &         2    &         5    &         1    &        8 \\
2003~AZ$_{84}$ &    x     &       18   &      18     &        19   &        16    &         9    &        15    &       11 \\
\hline
\end{tabular}
\begin{list}{}
\item Number of predicted events per year for each TNO.
\end{list}
\end{table}

\section{Discussion}

   We have presented predictions for stellar occultations by the large TNOs Eris, Haumea, Ixion, Makemake, Orcus, Quaoar, Sedna, Varuna, 2002~TX$_{300}$, and 2003~AZ$_{84}$ for 2008/2009--2015. We applied the same observational and astrometric procedures as a previous study dedicated to the Pluto system and published in \cite{Assafin2010}. Many issues were addressed in that work, such as body ephemeris offset determination, catalog zero-point position errors and field-of-view size, long-term predictions and stellar proper motions, and faint-visual versus bright infrared stars, which also hold true here.

   The local astrometric catalogs that we obtain are complete to about R = 19 and in the UCAC2 frame. Up to R magnitude completeness brightness, stars have positional errors of 40 mas or smaller and computed proper motions. The catalogs cover the TNO sky paths with $30\arcmin$ width in declination and can be very useful in astrometric follow-up programs on telescopes/CCDs with FOVs of any size. The catalogs are also sources of photometric information for calibration and occultation stars. The astrometry was made by the use of the PRAIA package (\cite{PRAIA}). About 316 GB of science-alone WFI images were treated, not to mention calibration observations to derive field distorion patterns, which improved the astrometric results of WFI data to the 40 mas level (see Sect. 9 in P1 for a detailed discussion about the field distortion pattern improvements for WFI astrometry).

   As an important by-product, the high-resolution CCD frames obtained  (1 pixel = 0$\farcs$238) form an image bank useful for the visual inspection of the fields of view. For instance, the analysis of small details in the calibration stars, close-by faint objects, and other elements in the images may prevent problems in the recording and photometry of occultation observations. When it comes to an event, charts extracted from this image bank are usually posted in our web page, but any particular field around some desired position can be obtained in FITS format on request to the authors. Public access to the original images from the observed WFI CCD mosaics will soon be available through the ESO Image Archive Portal.

    No threshold in R magnitude was established in the search for occultations, as faint R objects may turn out to be bright IR stars, suitable for observation at adequately equipped IR ground-based facilities or by the SOFIA observatory (\cite{Gehrz2009}); H, J, and K magnitudes are readily available in the catalog if the star belongs to the 2MASS. In addition, events that occur at slow shadow speeds of less than 20 km s$^{-1}$ may become observable. No constraints on a geographic place were applied, as in principle SOFIA observations can be done from any sub-solar point on Earth. Events in daylight at sub-planet point were not excluded either, as they remained observable in the dark, right above the horizon, from places near the Earth terminator. Throughout this paper, we do not distinguish between past and future predictions, publishing all found occultations for the sky path covered (or to be covered) by the TNOs over the years 2008/2009--2015. We assume that the WFI predictions of past occultations may be useful as either reference for ongoing fittings of recently obtained light curves, an aid to deriving ephemeris offsets by comparing expected and observed central instants and C/A values, or an external check for the accuracy and precision of our WFI predictions.

    The information in the prediction tables allows us to determine the central instant and shadow path over the Earth. For ground-based observations, the selection of events usually favors brighter stars and shadow paths that cross easily accessible places for the use of mobile telescopes. Faint stars may be selected in the case of observation at sites with larger observatories/telescopes. If the star is reddish, observations can be made in daylight with IR detectors. For TNOs with a low rate of occultations per
year, such as Eris, opportunities are very rare, so broader nets of observers may be formed over larger areas, even if the event is not as promising in terms of star brightness or location. We leave the choice of potential events among the many predicted occultations listed in the tables to the judgement of the reader. Our group publishes at our web page potential occultation campaigns on a yearly basis.

    The errors in our TNO sky path predictions are governed by the internal offset uncertainties listed in Table~\ref{table:7}. These uncertainties are based on optical observations made for a period of some years. The budget ephemeris error estimates given by JPL - which increase with each passing year - only express the random error features of the JPL orbit solutions, rather than represent a possible increase in systematic trends. As time goes by and more TNO observations and occultations are accumulated, more accurate JPL ephemeris     solutions are being published, although in the new versions no significant shifts in the TNO orbital sky paths are predicted for the next few years. For now, the adoption of a constant ephemeris offset in the predictions has the only side effect of absorbing some remaining systematic errors into the offset uncertainties. These uncertainties, on the other hand, do not increase with time for the period considered and are realistic estimates for the overall sky path prediction uncertainties. In the prediction tables (available in eletronic form), in addition to the star position, we indicate the estimated TNO position (with ephemeris offset applied) at the predicted occultation instant. This should help us to evaluate the events in the future as updates become available at the star position and on the TNO ephemeris by follow-up astrometric programs or by the JPL.

    A tentative way to estimate the probability of success in recording an occultation of a particular TNO is to consider the ratio of the apparent TNO diameter in mas to its ephemeris offset errors displayed in Table~\ref{table:7}. We list TNO diameter values in Table~\ref{table:10} to support reliability estimates for the predictions. They are given in km and as geocentric apparent sizes in mas for 2012.5. Values were obtained from previous stellar occultations and by modeling visible and IR observations (see references in Table~\ref{table:10}).

\begin{table}
\caption{TNO diameters.}
\label{table:10}
\centering
\begin{tabular}{c c c c}
\hline\hline
              & \multicolumn{3}{c}{TNO diameter}  \\
              &  km & mas & Reference \\
\hline                           
Eris           &   2326 ($+$/$-$12)     & 33 & \cite{Sicardy2011b} \\
Haumea         &   1324 ($+$/$-$167)    & 36 & \cite{Lellouch2010} \\
Ixion          &   0650 ($+$250/$-$220) & 22 & \cite{Stansberry2008} \\
Makemake       &   1455 ($+$/$-$12)     & 38 & \cite{Ortiz2011} \\
Orcus          &   0850 ($+$/$-$90)     & 25 & \cite{Lim2010} \\
Quaoar         &   1170 ($+$/$-$40)     & 34 & \cite{BragaRibas2011b} \\
Sedna          &   1200--1600           & 22 & \cite{Brown2008} \\
Varuna         &   1003 ($+$/$-$09)     & 32 & \cite{Sicardy2010} \\
2002~TX$_{300}$&   0143 ($+$/$-$05)     & 05 & \cite{Elliot2009} \\
2003~AZ$_{84}$ &   0910 ($+$/$-$60)     & 28 & \cite{Mueller2010} \\
\hline
\end{tabular}
\begin{list}{}
\item TNO diameters in km (uncertainties in parentheses) and as geocentric apparent sizes in mas for 2012.5. Values were obtained from previous stellar occultations and by modeling visible and IR observations (see references).
\end{list}
\end{table}

   The ESO2p2/WFI predictions can be directly compared to actually observed stellar occultations in order to estimate the external accuracy and precision of predictions. Table~\ref{table:11} displays a direct comparison between predictions (ESO2p2/WFI-based) and actually observed occultations, for which P/A, central instant and C/A values were obtained from fits to the light curves of the chords. We used six out of the seven TNO stellar occultations known to date, for which data is available (\cite{Elliot2009}; \cite{Sicardy2011b}; Braga-Ribas, 2011, priv. comm.). Using P/As, distances and relative speeds, we computed both the predicted minus fitted central instant and C/A differences. Differences in central instants are expressed in seconds and in mas (in order to be comparable with C/A results), and C/A differences are given in both mas and kilometers. We give in Table~\ref{table:11} the averages for the modulus of the differences. We note that the WFI prediction for the Eris' central instant was in strong disagreement with observations. This may indicate that much more accurate right ascension ephemeris offsets should be obtained in our follow-up program for Eris. A separate set of averages is provided in Table~\ref{table:11} without taking Eris into account. The TNO 2003~AZ$_{84}$ is excluded from the analysis as the fit to the actual occultation is based on the WFI star position, and the current WFI predictions already uses the offsets derived from the occultation fitting. In all, the differences in C/A and central instant are consistent with the expected errors in the WFI star positions as a function of magnitude and ephemeris offset error (see Table~\ref{table:7}). The somewhat larger differences found in Table~\ref{table:11} for the central instant may be associated with ephemeris offset uncertainties in right ascension (the direction of the chords are usually east-west). One possible cause is geocentric parallax error (see discussion about the follow-up program and ephemeris offset determination in the next paragraphs). In all, for R = 19 stars (catalog magnitude completeness) and 40 mas errors in the WFI positions, we may assume a bulk error of about 80 mas for C/A, that is dominated by the ephemeris offsets errors of about 70 mas. For about 40 A.U., this implies a shadow path uncertainty over the Earth of the order of 2300 km. If the ephemeris offsets can be well-determined to a more accurate precision than 30 mas, then a bulk error of 50 mas in C/A can be achieved, leading to a precision of about 1400 km for the WFI occultation path predictions, which is similar to the obtained values displayed in Table~\ref{table:11}.

\begin{table}
\caption{Comparison between WFI-based and actual observed occultations: central instants and C/As}
\label{table:11}
\centering
\begin{tabular}{l c c c c c}
\hline\hline
                     & Star    &\multicolumn{4}{c}{Prediction minus Occultation} \\
     TNO             & mag (R) &\multicolumn{2}{c}{Central instant}  &  \multicolumn{2}{c}{C/A}        \\
                     &         &  s          &     mas   &     mas   &  km        \\
\hline                           
2002~TX$_{300}$      &  12.9   &    $-$15    &   $-$13   &  $-$09    &   $-$265   \\
Varuna               &  11.0   &    $-$20    &   $-$12   &  $-$63    &   $-$1953  \\
Eris                 &  16.9   &    $-$604   &   $-$229  &  $+$39    &   $+$2708  \\
2003~AZ$_{84}$       &  18.2   &    $+$0     &   $+$0    &  $+$0     &   $+$0     \\
Makemake             &  18.4   &    $+$196   &   $+$116  &  $+$40    &   $+$1494  \\
Quaoar               &  15.7   &    $+$130   &   $+$78   &  $-$44    &   $-$1351  \\
average (moduli)     &         &  193        &   90      &  39       &   1554     \\
average (Eris out)   &         &  90         &   55      &  39       &   1266     \\
\hline
\end{tabular}
\begin{list}{}
\item  Comparison between central instants and C/As from WFI-based (Table~\ref{table:8}) and actual observations (fit to data) for 6 (out of seven) successfully recorded TNO stellar occultations to date for which data is available. Using P/A, distance and the relative speed, central instant differences are expressed in seconds and in mas, and C/A differences in mas and in km. Averages for the modulus of the differences are computed with and without Eris. 2003~AZ$_{84}$ is excluded from the analysis as the fit to the actual occultation already uses the WFI star position, and the current WFI predictions already uses the offsets derived from the occultation fitting.
\end{list}
\end{table}

   Long-term predictions are the only way of spotting potential stellar occultations in terms of favourable geographic access, observatories and instrumental facilities, etc. Once an event is chosen, then a complex international campaign is initiated, with astrometric follow-up of the TNO and star being mandatory. The only reliable long-term predictions of stellar occultations that use solely all-sky star catalogs are those based on the UCAC2. This limits the search for candidates to about R = 16 stars. The goal of our WFI program was to extend the UCAC2 position precision to stars as faint as about R = 19. The availability of data for such faint stars with about 40 mas position precision improves the chances of finding more suitable candidates for TNO occultations, owing to the increase in the number of suitable stars in the sky path and, consequently, the frequency of appulses. Owing to the recent developments of solid state detectors, and because TNOs are usually much fainter than the occulted stars, we can obtain good contrast in the light curves, even using very modest instruments. This has been verified for R = 17 -- 18 stars in the occultations by Eris and 2003 AZ$_{84}$ (\cite{Maury2010}; \cite{Sicardy2011b}; \cite{BragaRibas2011a}; Braga-Ribas et al. 2011, in prep.), observed with 40 cm and 60 cm telescopes in Chile.

    The sky paths covered by the star catalogs were determined by the chosen time span of our predictions (2008--2015) in a compromise between the availability of WFI telescope time and the number of selected TNOs. The upper limit coincides with NASA's New Horizons spacecraft arrival at Pluto in 2015. Given this long time coverage of about eight years, the computation of proper motions for the stars was also of concern. That is why we used as first epoch data the best available astrometric resources at the time of completion of this work. In all, the comparison made with actual events (Table~\ref{table:11}) and the successful WFI predictions specifically for the faint stars of Eris, 2003 AZ$_{84}$ and Makemake occultations indicate that the WFI astrometric solution is sactisfatory in the context of long-term predictions.

     Owing to the lack of observations of trans-neptunian objects for ephemerides, to the few positions sometimes derived without the appropriate astrometric care, and to the relatively short orbital arc length coverage, the actual position of TNOs can differ significantly from their ephemerides. To refine the orbits of large TNOs in the short and long runs, by regularly obtaining precise astrometric positions, a parallel program involving many telescopes of 60 cm to 2 m class in Chile, France, and Brazil was started in 2007 (see further details in Sect. 5). Using the first results of this program for the predictions published in this work, we applied average fixed offsets to the current JPL ephemerides of the TNOs in the list (Table~\ref{table:7}). In some cases, offsets reached almost 200 mas (about one Earth radius at 30 AU) or more.

     Given the time span of 2008--2015, we expect that the fixed corrections applied to the JPL ephemerides suffice for pinpointing the most interesting occultations among all predictions. However, once an event is selected, the shadow path on Earth must be further refined. From the point of view of TNO orbits, improvements can still be made to the accuracy of the WFI predictions. For instance, by observing TNOs at their opposition and conjunctions to obtain the observed ephemeris offsets, one can model the geocentric parallax errors and correctly extrapolate the observed ephemeris offsets with time for the occultation dates. For a body at 30 AU in the Ecliptic, these errors can amount to 100 mas in right ascension for an ephemeris error of some few tens of thousand kilometers in heliocentric distance (\cite{Camargo2010}). As successful occultations are recorded, the ephemeris offsets are more accurately determined. For three TNOs (see Table~\ref{table:7}), we could already improve the ephemeris offsets for this work using occultation results. As the new occultation results are consolidated, we will update the ephemeris offsets and release new, higher quality predictions.  

    An advancement that is beyond the scope of WFI predictions, for selected occultations, is an alternative procedure for improving the shadow path in the short term by direct fitting an orbital arc to the observed TNO positions. This can be done by numerically integrating orbits or, in principle, by using the current ephemeris as a template and then fitting it to the observations following a correction model. In the case of observations when both star and TNO are imaged in the same field of view, this procedure can yield very accurate predictions, as high precision relative positions can be attained for the TNO with respect to the occultation star, once proper care is taken to correct for differential color refraction.

   The astrometric follow-up of the selected stars and TNOs on a regular basis is therefore very important, regardless of the approach taken to improve the accuracy of the predictions. In practice, however, the week or two before occultation date is still decisive in pinning down the shadow path on Earth. Updates on the TNO ephemeris offsets, star positions of selected occultations, updated reports, and finding charts are available on a continuous basis by our group at our web page\footnote{http://www.lesia.obspm.fr/perso/bruno-sicardy/}. Among other similar projects conducted by other active groups in the U.S.A., our efforts reflect a long-term collaboration with the international campaigns coordinated by the Observatoire de Paris for this purpose.

\begin{acknowledgements}

M.A., J.I.B.C., and R.V.M. acknowledge CNPq grants 482080/2009-4, 306028/2005-0, 478318/2007-3, 151392/2005-6 and 304124/2007-9. M.A. and J.I.B.C. thank FAPERJ for grants E-26/170.686/2004 and E-26/110.177/2009. F.B.R thanks the financial support of CDFB/CAPES. The authors acknowledge J. Giorgini (JPL) for his help in the use of NAIF tools. The authors are grateful to F. Colas and F. Vachier for the kind access to unpublished TNO observations and/or positions from a number of telescopes, which in part contributed to the determination of the provisional TNO ephemeris offsets used in this work.

\end{acknowledgements}

\end{document}